# Imaging charge and spin diffusion of minority carriers in GaAs

I. Favorskiy[1], D. Vu[1], E. Peytavit[2], S. Arscott[2], D. Paget[1] and A. C. H. Rowe[1]

[1] Physique de la Matière Condensée, Ecole Polytechnique CNRS, 91128 Palaiseau, France
[2] Institut d'Electronique, de Microélectronique et de Nanotechnologie (IEMN), University of Lille, CNRS, Avenue Poincaré, Cité Scientifique, 59652 Villeneuve d'Ascq, France

**Abstract**

Room temperature electronic diffusion is studied in 3 μm thick epitaxial p+ GaAs lift-off films using a novel circularly polarized photoluminescence microscope. The method is equivalent to using a standard optical microscope and provides a contactless means to measure charge ($L$) and spin ($L_s$) diffusion lengths. The measured values of $L$ and $L_s$ are in excellent agreement with the spatially averaged polarization and a sharp reduction in these two quantities ($L$ from 21.3 μm to 1.2 μm and $L_s$ from 1.3 μm to 0.8 μm) is measured with increasing surface recombination. Outwards diffusion results in a factor of 10 increase in the polarization at the excitation spot.

**Text**

Understanding spin transport in semiconductors, particularly in GaAs[1] and more recently in silicon,[2] is potentially important for future spintronic applications.[3] While recent effort has been directed at measuring and describing new phenomena such as a relative reduction in the spin diffusion coefficient $(D_s)$[4] and unusual spin-orbit interaction induced effects along selected crystal axes,[5] a recurring theme in the literature is the measurement of the characteristic length over which the electronic spin polarization can be transported. In non centro-symmetric semiconductors such as GaAs where the D'yakanov-Perel mechanism limits spin lifetime $(\tau_s)$, spin can be transported under the effect of an applied electric field over more than 100 μm at cryogenic temperatures.[6] In the absence of electric fields, spin diffusion lengths $(L_s)$ of several μm have been reported, mainly in n-type GaAs,[7] while at room temperature more modest values of the order of 1 μm are expected. In centro-symmetric materials where the spin-orbit interaction is weak (and thus $\tau_s$ long) such as silicon, much larger room temperature spin drift and diffusion lengths are obtained.[8]

Apart from non local spin accumulation experiments requiring several samples with electrical contacts spaced by different lengths,[9] $L_s$ (but not $L$) is typically measured using spatially resolved Faraday or Kerr rotation (also known as Kerr microscopy).[10] When combined with time resolution, a direct measure of $D_s = L_s^2/\tau_s$ is possible.[11] $D_s$ is also measured in four-wave spin grating experiments in which the charge diffusion coefficient $(D_e)$ is also measured.[12] Although the spin grating technique is powerful, it is experimentally complex and difficult to master. Here we use a novel circularly polarized photo-luminescence (PL) microscopy to study both lateral charge and spin diffusion in GaAs. The microscope is a modified Nikon Optiphot 70 with λ/4 plates at the excitation and reception that permit a spatially resolved analysis of the circularly polarized PL. A beam splitter at the reception also enables spatially resolved PL spectra to be measured with a fiber coupled spectrometer. Being a far field technique, spatial resolution is limited by diffraction and is thus comparable with that achieved in Kerr microscopy. Although the spatial resolution of the spin grating method is superior, the technique used here is far simpler, and in practice is equivalent to taking images on a standard laboratory optical microscope. A similar technique was recently used to measure the minority carrier charge diffusion length ($L$) in GaAs.[13] As with Kerr microscopy, when combined with time resolution, a direct measurement of minority carrier lifetime ($\tau$) and $D_e$ is possible.[14]

The samples studied here are 3 $\mu$m thick $p+$ doped GaAs thin film patches collected using an epitaxial liftoff technique and assembled onto SiC substrates.[15] The lateral dimensions of each patch, 400 $\mu$m x 400 $\mu$m, are sufficiently large that edge effects can be neglected. In one case ($N_A = 10^{18}$ cm$^{-3}$) the two surfaces are covered with the native oxide and in the other ($N_A = 1.5$ x $10^{17}$ cm$^{-3}$) they are terminated with 50 nm thick layers of $Ga_{0.51}In_{0.49}P$. The main effect of the different surface terminations is to modify the surface recombination velocity from $S = 10^7$ cms$^{-1}$ for the native oxide covered sample[16] to approximately $S = 10^3$ cms$^{-1}$ for the "passivated" sample.[17] Note that since vertical diffusion over a distance larger than the effective depth of field of the microscope objective results in a defocused luminescence spot, this may induce systematic errors in the estimated diffusion length. It is therefore important to use thin layers of thickness $d \leq fn$ where $f$ is the objective depth of field ($\sim 0.4$ $\mu$m) and $n$ is the refractive index of the semiconductor.

The samples are excited with a 50 $\mu$W circularly polarized laser at 1.59 eV focussed through a x100 microscope objective to a Gaussian spot of half width, $w = 0.9$ $\mu$m (see inset, Fig. 1 for a white light image of a patch with laser spot). In order to negate residual birefringence in the optical path, both $\sigma^+$ and $\sigma^-$ polarized components are in turn used to excite the sample and an image is taken of the $\sigma^{\pm}$ polarized PL with the laser being removed by an appropriate filter. The resulting four images, denoted $\sigma^{++}$, $\sigma^{+-}$, $\sigma^{--}$ and $\sigma^{-+}$ are combined to form a sum image ($I_s$ = [$\sigma^{++}$ + $\sigma^{+-}$ + $\sigma^{--}$ + $\sigma^{-+}$]/2) and a difference image ($I_d$ = [$\sigma^{++}$ - $\sigma^{+-}$ + $\sigma^{--}$ - $\sigma^{-+}$]/2). The spectra obtained at the center of these images are shown for the passivated sample in Fig. 1 while the images themselves are shown in Figs. 2a and 2b. Provided the concentration of photoelectrons is much smaller than $N_A$ one has[18] $I_s = K(n_+ + n_-)$ and $I_d = KP_i(n_+ - n_-)$, where $n_+ + n_-$ and $n_+ - n_-$ are the sum and difference concentrations of photoelectrons of spin $\pm$ along the direction $z$ of light excitation. For GaAs excited close to its bandgap the quantity $|P_i|$ depends on the matrix elements for recombination and is equal to 0.5. The constant $K$ only depends on the ratios of radiative to nonradiative lifetimes and on the parameters of the PL detection.

The spectra in Fig. 1 are obtained via a 50 μm diameter optical fiber whose diameter corresponds to a spot of diameter $l \sim 1$ μm on the images of Fig. 2. The residual integrated laser intensity at the centre of the PL image is almost a factor of 1000 weaker than the PL intensity and drops off even further at large distances from the spot center. Since the laser is strongly polarized, a negligible component appears in the difference spectrum. The ratio of the integrated difference to sum spectra between 1.4 eV and 1.5 eV yields a center polarization of $P_c \sim 3.4$ % in reasonable agreement with that obtained at the centers of the images in Figs. 2a and 2b ($\sim$ 2 %). In addition to confirming that the images in Fig. 2 contain no significant contribution from the laser, the spectrometer also shows that local sample heating is negligible.

The steady-state diffusion equations for $n_+ + n_-$ and $n_+ - n_-$ are respectively

$$0 = (g_+ + g_-)\tau\varphi(r)\alpha e^{-\alpha z} - (n_+ + n_-) + L^2\Delta(n_+ + n_-) \tag{1a}$$

$$0 = (g_+ - g_-)\tau_s\varphi(r)\alpha e^{-\alpha z} - (n_+ - n_-) + L_s^2\Delta(n_+ - n_-) \tag{1b}$$

where $\alpha = 1$ μm$^{-1}$ is the coefficient of light absorption along the normal $z$ to the surface and $\varphi(r)$ describes the radial (Gaussian) dependence of the light excitation intensity. The creation rates $g_\pm$ for electrons of $\pm$ spin are of the form $g(1\pm P_i)$ for σ$^-$ polarized excitation and $g(1\mp P_i)$ for σ$^+$ polarized excitation. Neglecting a small possible difference in $D_e$ and $D_s$,[4] these equations give the spatial dependence of $n_+ \pm n_-$ once the boundary conditions

$$\pm\frac{\partial(n_+ \pm n_-)}{\partial z} = \frac{S}{D_i}(n_+ \pm n_-)$$ have been applied at the top (+) and bottom (-) faces (with $D_i = D_e = D_s$).

Figures 2c and 2d show respectively the angular averaged profiles (open circles) of the decay of $I_s$ and $I_d$ for the passivated sample with $r$ measured from the center of the PL spot. In the case of $I_s$ the spot extends over more than 100 μm laterally, far larger than $w$ (see dotted line for laser profile), and is the result of diffusion of photoelectrons described by Eq. (1a). If $L \gg d$ the diffusion is 2D and if, in addition, $S < 2D_e/d \sim 10^5$ cms$^{-1}$, Eq. (1a) has an analytic solution in the form of a Bessel function, K$_0(r/L)$.[19] A convolution of K$_0(r/L)$ with $\varphi(r)$ can be fitted to the experimental curve in Fig. 2c where the only fitting parameter is $L$. For $L = 21.3$ μm, the solid black curve in Fig. 2c is obtained. Note that $L \gg d$ so the assumption of 2D diffusion is valid, but that the data only extend out to about $r = 50$ μm, insufficient for taking an exponential approximation to K$_0(r/L)$.[13] Since $S < 2D_e/d$ the value obtained for $L = \sqrt{D_e\tau}$ is not modified by the weak surface recombination since and corresponds to the *intrinsic* charge diffusion length.

For $I_d$ (Fig. 2b) the decay with $r$ is far more rapid, reflecting the fact that $\tau_s \ll \tau$ and thus $L_s = \sqrt{D_s \tau} \ll L$. The lateral extent of the laser (dotted curve) is still inferior to that of the difference signal meaning that the laser spot size does not limit the measurement of $L_s$. It is no longer possible to fit a convolved Bessel function to this data since the lateral extent of the spot is comparable to $d$ and the diffusion is therefore only quasi-2D. Instead solutions to Eq. (1b) must be found numerically. This is achieved here using a commercial finite element package. For $L_s = 1.2$ μm, the solid black curve in Fig. 2d is obtained in excellent agreement with the data. Again, since $\tau_s \sim 100$ ps[20] this is the *intrinsic* spin diffusion length.

The spatially averaged polarization, $P$, obtained from the ratio of the integrated intensities of $I_d$ to $I_s$ is[18] $P = P_i \dfrac{\tau_s}{\tau} = 0.16\,\%$. Here a slight depolarization of the laser on transit through the microscope objective is accounted for ($P_{las} = 70\,\%$). Neglecting a small possible difference in $D_s$ and $D_e$ due to spin drag,[4] $P$ should also be equal to $P_i\left(\dfrac{L_s}{L}\right)^2$ which, using the intrinsic values obtained for $L$ and $L_s$ yields 0.11 %. The diffusion lengths thus appear in very good agreement with the polarization which would be obtained in a standard optical pumping experiment on this sample. It is interesting to note that $P_c$ is much larger than $P$. This is the result of outwards carrier diffusion that reduces the residence time at the spot centre (i.e. the lifetime at the center) of typical dimension $l$ to $1/\tau^* = 1/\tau + D_e/l^2$. Since $\tau \sim 10$ ns and $D_e \sim 2.5$ x $10^9$ μm$^2$/s for GaAs of this doping level,[21] $\tau^* \sim \tau/30$ and consequently $P_c \sim 30P$ as observed.

The same experiment is carried out on the naturally oxidized sample, with the angular averaged profiles for $I_s$ and $I_d$ shown in Figs. 3a and 3b respectively (open circles). In both cases the lateral extent of the PL is now comparable with $d$ but still larger than $w$ (dotted line).

Thus numerical modelling of Eqs. (1a) and (1b) is again required as both the charge and spin diffusion are quasi-2D. Surface recombination can be accounted for in one of two ways. Usually, the aforementioned boundary conditions are applied with $S = 10^7$ cms$^{-1}$ for native oxide covered GaAs[16] and with $L$ and $L_s$ fixed at the intrinsic values measured on the passivated sample. Doing this yields the solid curve with closed circles in Figs. 3a and 3b which closely resemble the data. Alternatively the boundary conditions can be applied with $S = 10^3$ cms$^{-1}$ (as for the naturally oxidized sample) and the increased surface recombination accounted for via a reduced effective lifetime. This procedure results $L_{eff} = 1.3$ μm and $L_{s,eff} = 0.8$ μm (see solid curves in Figs. 3a and 3b). With these values $P = 15$ % is expected, which compares favorably with the ratio of the integrated intensities of $I_d$ to $I_s$ (12.1 %). Moreover, despite the diffusion being only quasi-2D, these values are in good agreement with the 2D diffusion result,[22] $\dfrac{1}{\tau_{eff}} = \dfrac{d^2}{\pi^2 D_e} + \dfrac{d}{2S} \approx \dfrac{d^2}{\pi^2 D_e}$ which yields an effective diffusion length $L_{eff} = \sqrt{D_e \tau_{eff}} = d/\pi \sim 0.95$ μm. This clearly demonstrates that surface recombination is primarily responsible for the sharp reductions in both $L$ and $L_s$ and that the difference in doping densities of the two samples plays only a secondary role.[20]

In most proposed bipolar spintronic devices,[15,23,24,25] $L_s$ is a parameter affecting device performance, so the method described here may be a useful for determining design constraints for such devices. Furthermore, a joint measurement of $D_e$ and $D_s$ is possible either by a time resolved measurement or by an in-situ study of the precession due to an applied transverse magnetic field.[20]

**Acknowledgements**

The authors acknowledge T. Verdier and T. Porteboeuf for their help in setting up the circularly polarized optical microscope, J-F. Lampin for fruitful discussions and X. Wallart for the epitaxial growth. This work partially funded by the ANR SPINJECT-06-BLAN-0253.

**Figures**

Fig. 1: Sum (open circles) and difference (closed circles) spectra obtained on the passivated

sample at the center of the PL images. The inset shows a white light image of the GaAs patch with the laser spot close to the center.

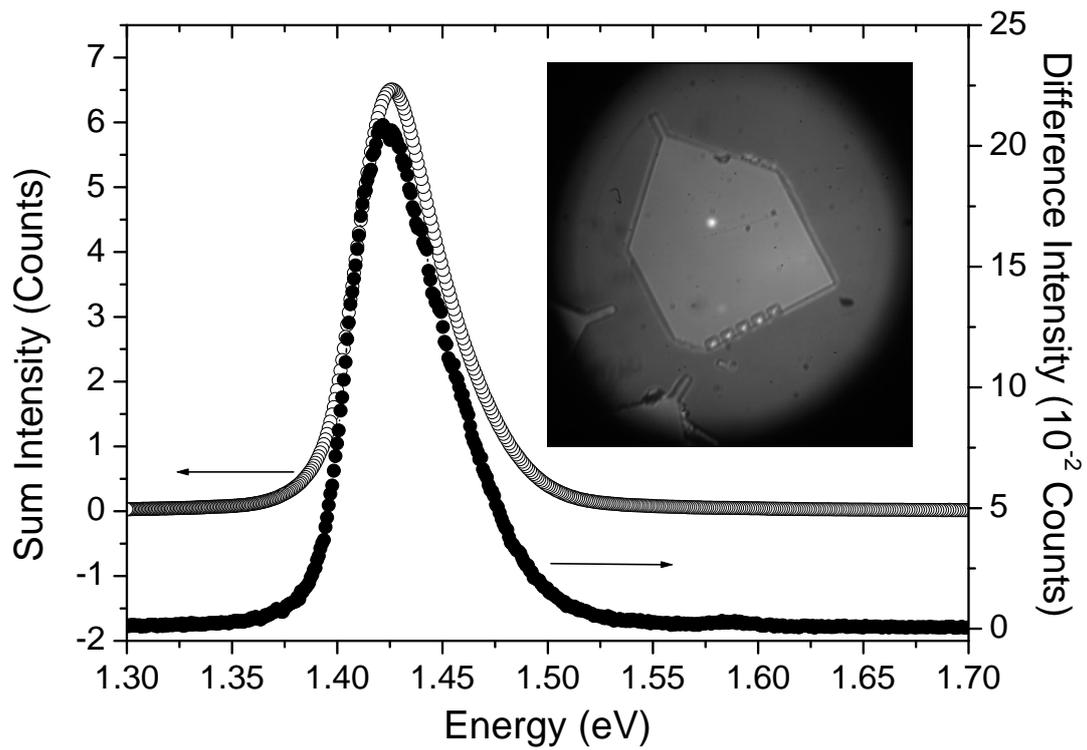

Fig. 2: The sum (a) and difference (b) images obtained on the passivated sample. (c) The angular average profile of $I_s$ plotted against $r$ (open circles). The laser excitation profile (dotted line) is smaller than the profile and so does not limit the measurement of $L$ and $L_s$. The solid line is obtained by convolving the excitation function with a Bessel function $K_0(r/L)$ with $L = 21.3$ μm. (d) $L_s$ is obtained by fitting the angular averaged profile of $I_d$ (open circles) with a numerical solution of Eq. (1b) with $L_s = 1.2$ μm.

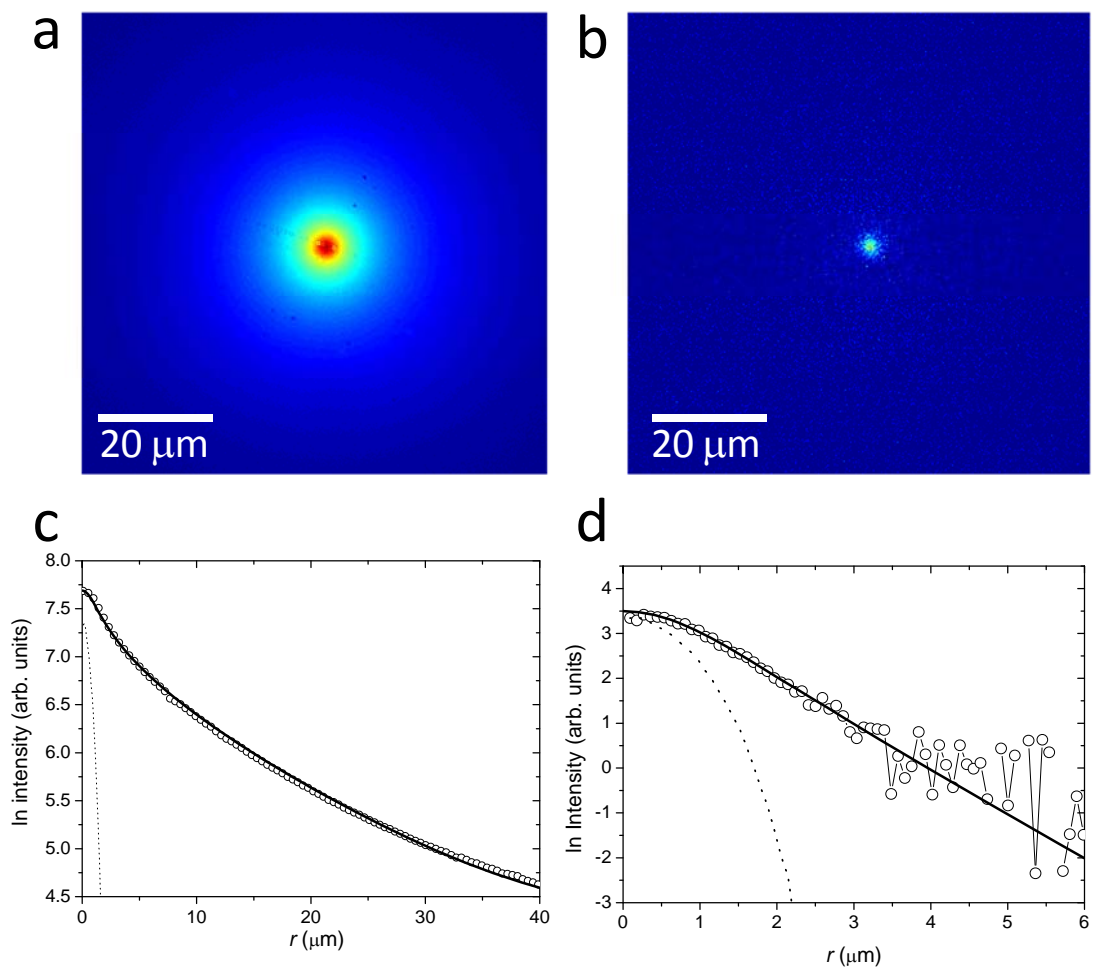

Fig. 3: (a) Open circles: Angular averaged profile of $I_s$ obtained on the naturally oxidized sample. Solid line: numerical resolution of Eq. (1a) with $S = 10^3$ cm/s and $L_{eff} = 1.3$ μm. Closed circles: numerical resolution for $S = 10^7$ cm/s and $L = 21.3$ μm. (b) Shows the same three curves for $I_d$ with $L_{s,eff} = 0.8$ μm and $L_s = 1.2$ μm. In both cases the lateral extent of the laser (dotted lines) is smaller than that of the luminescence profile.

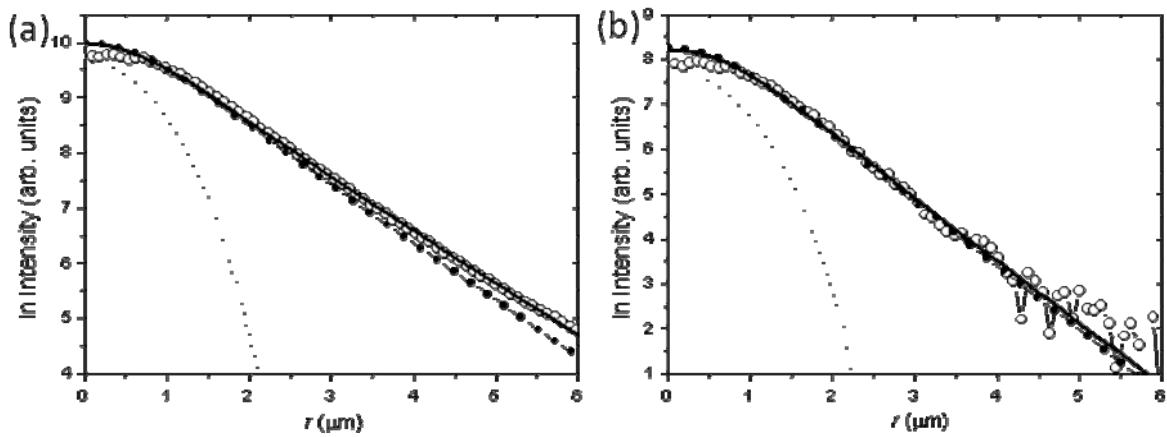